# Can Large Language Models Evaluate Grant Proposal Quality? Revisiting the Wennerås and Wold Peer Review Data

Ulf Sandström
Industrial Economics and Management, KTH Royal Institute of Technology, Stockholm, Sweden. https://orcid.org/0000-0003-1292-8239 ulf.sandstrom@indek.kth.se, forskningspolitik@gmail.com

Mike Thelwall
School of Information, Journalism and Communication, University of Sheffield, UK. https://orcid.org/0000-0001-6065-205X m.a.thelwall@sheffield.ac.uk

**Purpose**: Despite the importance of peer review for grant funding decisions, academics are often reluctant to conduct it. This can lead to long delays between submission and the final decision as well as the risk of substandard reviews from busy or non-specialist scholars. At least one funder now uses Large Language Models (LLMs) to reduce the reviewing burden but the accuracy of LLMs for scoring grant proposals needs to be assessed.
**Design/methodology/approach**: This article compares scores from a range of medium sized open weights LLMs with peer review scores for a well-researched dataset, the Swedish Medical Council's post-doctoral fellowship applications from 1994.
**Findings**: Whilst the LLM scores correlate moderately between each other (mean Spearman correlation: 0.34), they correlated weakly but positively and mostly statistically significantly with the average expert scores (mean Spearman correlation: 0.22). The highest rank correlation between expert scores and LLMs was 0.33 for Gemma 3 27b based on proposal titles and summaries without their main texts, which is about half (56%) of the correlation between reviewers.
**Research limitations**: The small sample size, old funding call and heterogeneous evaluation criteria all undermine the robustness of the analysis.
**Practical implications**: Despite the ability of LLMs to score grant proposals being quantitatively weaker than that of experts, at least in this special case, they may have role in application triage or tie-breaking.
**Originality/value**: This is the first assessment of the value of LLM scores for funding proposals.
**Keywords**: Grant peer review; Large Language Models; Open Weights LLMs; Scientometrics; Research evaluation

## 1 Introduction

Competitive research funding is a central part of most research systems. It is a process to direct money to where it can be used effectively, usually in support of loosely or tightly defined funder goals. The decision process typically involves peer review, usually with academics selected for relevant expertise and invited to submit an evaluation with a score, and sometimes several scores. A panel may then deliberate for a final decision based on these reports and their own judgement. Decisions can be slow because reviewers are reluctant (e.g., 9 out of 10 UK Research and Innovation reviewing requests are rejected: personal communication) and need to be given time to fit the review within potentially busy schedules. The key problems are that (a) the system consumes expert reviewer time, often unproductively for proposals that are rejected and with limited learning for experienced reviewers, and (b) the delays can degrade the value of the research and make it difficult to fund urgent or timely ideas.

Funders already use Artificial Intelligence (AI) to help select reviewers (Newman-Griffis et al., 2025) to improve efficiency and one now uses it for reviewing. La Caixa employs Large Language Models (LLMs) to triage proposals, with the lowest scoring proposals being sent to human experts for quick reviews. After this, proposals judged weak by all three (LLM and two reviewers) are rejected without a full review (three detailed expert reviews). This has led to a small reduction in the number of reviews needed (Carbonell Cortés et al., 2024).

In the above context and given that LLMs have shown a moderately good ability to estimate the research quality of published academic journal articles (Thelwall, 2025; Thelwall & Yang, 2025) and other academic documents, such as research environment statements (Kousha et al., 2025), conference papers (Thelwall & Yaghi, 2025a) and books (Thelwall & Cox, 2025), it is important to investigate whether they can evaluate the quality of grant funding applications. While some previous studies have used AI to assess funding applications (e.g., Sikimić, & Radovanović, 2022), none seem to have produced results that were good enough to be adopted by funders, so La Caixa seems to be the pioneer. Two practical complications for researchers are the fact that applications are usually confidential, unlike open access journal articles, so require additional steps and safeguards to access, and there are many types of funding scheme and peer review process. These problems are (partially) addressed in the current study by focusing on a set of funding proposals of one type, Swedish MRC post-doctoral fellowship applications from 1994; these are available due to the principle of public access to official documents in Sweden.[1]

The current study assesses the ability of LLMs to rate funding proposals in a way that aligns with expert scores. It focuses on Swedish medical fellowship applications from 1994 for the pragmatic reason that they are available and getting evidence for this case will be suggestive about how well LLMs will perform in other cases. It also focuses on medium sized "open weights" LLMs: these can be downloaded and run locally, which seems likely to be a requirement for funder grant proposal evaluation due to the sensitivity of the information. Previous research has suggested that these LLMs work nearly as well as larger cloud-based models such as ChatGPT for academic evaluations (Thelwall & Mohammadi, 2026). The following research questions, all concerning Swedish medical fellowship applications from 1994, drive the study.

1. Do LLM scores correlate positively with expert review scores? This seems necessary for LLMs to have any role in evaluating proposals.
2. Do scores from different LLMs correlate with each other? This will reveal the extent to which there is a common "LLM quality" that they all attempt to measure.
3. How does the level of agreement between LLMs and expert reviewers compare to the level of agreement between different expert reviewers? This is important to contextualise the answer to RQ1.

---

[1] The proposals were received from Agnes Wold; grant proposals to the MRC are available pursuant to the Swedish principle of public access to official documents, which grants the public a constitutional right to access official records held by public authorities (Freedom of the Press Act, Chapter 2).

# 2 Background

The empirical basis for this study consists of post-doctoral fellowship applications submitted to the Swedish Medical Research Council (MRC) in 1994. At the time, these fellowships represented some of the most prestigious and competitive early-career funding opportunities in Swedish biomedicine. They were aimed at recently defended PhDs and designed to support the transition to independent research, often involving a period of training in a new research environment. Applications were evaluated by disciplinary panels through a structured peer-review process that combined several criteria, typically including the applicant's prior achievements, the scientific quality and feasibility of the proposed project, and the suitability of the host environment. Reviewers provided numerical scores, which were then aggregated and used by panels as the basis for funding decisions. Crucially for the present study, both the applications and the associated review documentation, including the scores, are accessible due to Sweden's long-standing principle of public access to official records.

These same applications later formed the empirical foundation of the influential study by Wennerås and Wold (1997), which examined gender bias in peer review at the Swedish MRC. Their analysis revealed systematic disparities in reviewer scores that could not be accounted for by conventional productivity indicators alone, showing that female applicants had to be substantially more productive than male applicants to receive comparable scores. Beyond its substantive findings, the Wennerås and Wold paper became methodologically important because it linked quantitative peer-review scores to independent bibliometric measures of applicant merit. As a result, it has become one of the most cited and frequently discussed studies in the research evaluation literature (Sandström & Riis, 2025).

Subsequent reanalyses and discussions of the Wennerås and Wold material have drawn attention to additional features of the MRC review process, including the multidimensional nature of the assessment criteria and the role of panels in shaping and moderating individual reviewer judgements. Taken together, this body of work has established the 1994 MRC post-doctoral fellowship competition as a benchmark case for examining the functioning, strengths, and limitations of grant peer review. In particular, it offers rare insight into how numerical scores emerge from expert judgement under realistic funding conditions, rather than in experimental or simulated settings.

Against this background, the present study repurposes the same historical material to address a different but complementary question: whether contemporary Large Language Models can approximate expert peer-review scores when applied to real grant proposals. The value of this case lies not in its recency, but in the unusually rich contextual knowledge available about the review process and its outcomes. This makes it possible to interpret LLM–reviewer agreement not in isolation, but in relation to known levels of reviewer–reviewer agreement and previously documented sources of variation and bias in expert scoring.

## 2.1 Inter-reviewer reliability for grant peer review

A long-standing concern in research evaluation is the degree to which different reviewers agree when assessing the same grant proposal. Inter-reviewer reliability is typically modest in grant peer review, regardless of discipline, funding scheme, or country. Early quantitative studies already showed that agreement between reviewers is often low, with correlations between independent scores commonly ranging from weak to moderate (e.g., Cole et al, 1981; Cicchetti, 1991). Subsequent large-scale analyses have confirmed this pattern across biomedical, social science, and interdisciplinary funding contexts.

More recent evidence suggests that low inter-reviewer agreement is not primarily a consequence of reviewer incompetence but rather reflects the inherently multidimensional and judgement-based nature of grant assessment. Reviewers weigh criteria such as originality, feasibility, investigator track record, and potential impact differently, even when operating under shared guidelines (Langfeldt, 2006; Fogelholm et al., 2012). As a result, variability between reviewers is often systematic rather than random, with different reviewers emphasising different aspects of proposal quality.

Empirical studies of major funding agencies consistently find that inter-reviewer correlations for grant scores are substantially lower than those typically observed for journal peer review or bibliometric indicators. For example, analyses of NIH, NSF, and European funding programmes report correlations between reviewers that often fall in the range of 0.2–0.4, depending on field and design (Bornmann et al., 2010; Li & Agha, 2015). Panel discussions can reduce some of this dispersion, but they rarely eliminate it entirely, and final funding decisions often depend on negotiated judgements rather than purely aggregated scores.

This body of literature has important implications for studies comparing alternative evaluation methods to peer review. Low inter-reviewer reliability sets an upper bound on the level of agreement that any external system – including bibliometric indicators or Large Language Models – can reasonably be expected to achieve. Consequently, weak-to-moderate correlations between LLM-generated scores and expert reviewer scores should be interpreted considering the baseline level of disagreement among experts themselves, rather than as evidence of fundamental inadequacy.

## 2.2 LLMs for research evaluation

Many studies have investigated whether LLMs can score published journal articles for research quality or whether they can give useful feedback suggestions for authors during computing conference peer review. Both tasks are relevant here, although different from grant proposal scoring.

A range of cloud-based and open weights LLMs have an ability to score published journal articles for research quality in the sense of giving scores that correlate positively and statistically significantly with expert review scores (Thelwall, 2025; Thelwall & Mohammadi, 2026; Thelwall & Yang, 2025). The strength of the correlation varies between weak and moderate, depending on the field, with stronger correlations for health, life and physical sciences (Thelwall & Yaghi, 2025b). This correlation occurs despite the LLMs being fed with article titles and abstracts only, which is insufficient to evaluate an article. Thus, the LLM scores are just guesses, despite correlating positively with expert scores. The reason for the level of agreement is unknown but LLMs may identify topics or methods choices that expert reviewers tend to value above or below average in a field (e.g., Nunkoo & Thelwall, 2025).

In addition to estimating post-publication quality scores, LLMs can make suggestions about unpublished manuscripts from their full text as part of the peer review process (Liang et al., 2024) and can summarise reviewers' comments to support editors or authors (Du et al., 2024). Currently, the formal use of LLMs for peer review is supported only by a few computer science conferences and one publisher, but reviewers often elicit LLM help even when explicitly told not to (Naddaf, 2026), leading to the development of methods to detect LLM-authored reviews (Liu et al., 2025; Ryu et al., 2025).

The International Conference on Learning Representations (ICLR) is a venue that offers LLM suggestions to reviewers after completing their initial reviews. Of those accepting the feedback, 27% subsequently altered their reviews (Thakkar et al., 2025) so they seemed to be useful in this context. LLM suggestions might be particularly valued by new reviewers who are uncertain about what is appropriate for the task (Yun et al., 2026). Smaller open weights LLMs seem to be powerful enough for this task (Sadallah et al., 2025), allowing more security than when using cloud-based versions.

Agent-based LLM review support may be the future, such as with separate agents for writing, organising and filtering reviews (Bharti et al., 2026a; Farooq & Iqbal, 2026; Goyal et al., 2026). Completely banning LLMs from peer review seems unlikely to work, so ensuring their effective uses seems to be a more reasonable approach (e.g., Parra et al., 2026), especially in the context of the increasing burden of peer review.

# 3 Methods

The research design was to convert the Swedish medical fellowship applications from 1994 into English-language plain text documents and then submit them to a range of LLMs for scoring, then compare the scores with their expert scores. Because the expert peer-review scores were produced by multiple panels operating under different program mandates, particular care is required in how these scores are aggregated and interpreted; this issue is addressed explicitly below.

## *3.1 Data*

### 3.1.1 Applications

The Swedish medical fellowship applications are printouts of word-processed documents completed within a standard template. The sections include personal information about the applicant, a summary, and a description of the proposed work programme. One of the proposals only had a summary and was a "dummy" proposal in that the application had been approved as a double proposal in the previous funding round. This was removed. All proposals were written in English or Swedish. For processing by local LLMs, they were converted into plain text English language files as follows.

The 142 proposals were scanned into PDFs and then the summary and main proposal sections were extracted. These two sections, still in PDF format, were converted into plain text files using the text recognition software Tesseract (Smith, 2007), with the setting to extract both Swedish and English. This software is partly designed for poor quality scanned documents and for non-English European languages so it seemed appropriate for the task. This process worked generally well but with some mistakes and many errors with chemical formulae. Tesseract also attempted to convert the text within images, such as figure axis legends. All proposals were individually corrected for these anomalies by a native speaker (Ulf for Swedish, Mike for English).

The Swedish-language proposals and summaries were translated into English by the LLM Gemma 3 27b (see below), telling it to closely follow the structure of the original proposal. Without this instruction, it might have restructured proposals to make them more coherent or more natural for English. Such a restructuring could introduce errors due to misunderstandings or could improve on the applicant's communication skills so would not be desirable in this context. Translation is a core task for LLMs, and Gemma 3 27b seemed to perform it flawlessly. Nevertheless, all translated documents were checked by a bilingual native speaker of Swedish (Ulf) for translation accuracy. The following prompt was used.

Translate the following Swedish academic text into formal academic English text.

The result of this process was a set of 142 English-language proposal title and summary documents, and 142 title, summary, and proposal documents. Although it is more logical to assess a full proposal than just its summary, LLMs have been shown to work as well on summaries as with full texts for academic texts so both approaches were used. In both cases, only the proposal text was used, after redacting personal information. Figures were excluded because offline LLMs do not yet process them in documents. This omission is probably less critical than for publications because the figures usually offered visual explanations of the text rather than research results. In addition, applicant personal information (e.g., phone number, address in the cover page) was not processed.

### 3.1.2  Peer review scores

All proposals had been assessed and scored by independent reviewers, allocating a number in the range 0 to 4 for each of the following:
- Research question
- Methodology
- Applicant competence

Panel deliberations seemed to also incorporate the host research environment (Sandström, & Riis, 2025), but this was not an explicit criterion so was not included. Although all application forms were in the same format, candidates could submit them to one or more disciplinary panels. Thus, to some extent, the proposal texts target different goals, and sometimes multiple goals. Nevertheless, they all have the same broad aim in the sense of proposing a relevant innovative programme of medical research. Therefore, it makes sense to analyse them both as a single group and separately within panels, although the latter was not possible due to small sample sizes.

In addition to the panel variations, each application could be evaluated within up to three distinct funding programs - General (n=119), Priority Areas (n=19), and Mobility (n=24). Within each program, applications were assessed independently by up to two expert panels (denoted PK1 and PK2 in the data), each providing numerical scores on three dimensions: research question, methodology, and applicant competence. Consequently, a single applicant could receive multiple scores for the same dimension, originating from different panels and program contexts. A proposal might be sent to two panels if it was perceived to be interdisciplinary and relevant to both.

Because panel evaluations are nested within programs and because programs differ in mandate and evaluative criteria, there is no uniquely correct way to aggregate these scores into a single applicant-level measure. We treated each panel score as a distinct expert judgement and preserved all raw scores. We then aggregated scores in two explicit stages. In the first stage, panel scores were combined *within each program* (using the mean of PK1 and PK2, where both were available). We analysed all scores together and separately by *programme*.

Ideally, the analyses would have been repeated with different aggregation rules to avoid reliance on a single constructed score across heterogeneous evaluation contexts but the sample sizes were not large enough to make the results statistically distinguishable and so we only report the simplest possible approach. Unfortunately, averaging scores across programs implicitly assumes commensurability of evaluations produced under different mandates, which is not necessarily true and is a limitation.

## 3.2 LLM processing

### 3.2.1 Prompts

Prompts were generated to explain the task. The ideal prompts would mimic the instructions given to human reviewers. In this case, the instructions were not known and so the scale descriptions were used instead. Since the timeliness of an application can be critical, the submission year 1994 was included in the prompt repeatedly to reduce the chance that the proposals were downgraded for being out of date. After experiments with UK grant applications to refine the wording and ensure that the prompts worked, the text below was used.

> Analyse the strength of the following research proposal which was submitted to the Swedish Medical Research Council in 1994. Take into account the submission year 1994 when evaluating it. Finish your report with scores of 0-4 including fractions if necessary on the following scales:
> Research question: ___
> Where 0 = Incorrect; 1 = Unclear: the research question is not in line with the current research frontier. The study has been carried out previously or is being conducted along different lines; 2 = Clear and interesting; 3 = Clear and very interesting; 4 = Clear and original for 1994.
> Methodology: ___
> Where 0 = Incorrect; 1 = Material or method not fully appropriate; 2 = Material appropriate and method conventional but fit for purpose; 3 = Material appropriate and method appropriate and partly new; 4 = Original material or difficult-to-access material and/or technically original for 1994 and developed by the applicant.
> Applicant Competence: ___
> Where 0 = Missing; 1 = Questionable; 2 = Good; 3 = Very good; 4 = Outstandingly good for 1994.

### 3.2.2 LLMs

A range of modern medium sized LLMs was chosen for evaluation. Research funders seem unlikely to have the resources to run large models, such as the full DeepSeek, but medium sized models can run either quickly on specialised equipment or slowly on standard personal computers. They therefore seem to be at the edge of what is practical for a typical funder. There are many different LLMs, including older ones and academic custom versions of newer ones, but to give the most general results, only well-known recent LLMs were chosen. Both reasoning and non-reasoning models were included to allow a comparison of their abilities. The following LLMs were therefore chosen, all of which were the latest version of the family at the time of the experiments. The number before "b" refers to billions of parameters in the model and is an approximate indicator of how large they are.

- Gemma 3 27b: Non-reasoning model from Google.
- Magistral Small (24b). Reasoning model from Mistral AI.
- Llama Scout (17b). Reasoning model from Meta.

- Qwen3 (32b). Reasoning model from Alibaba.
- DeepSeek-R1 (32b). Reasoning model from DeepSeek.
- DeepSeek-R1 (72b). Large reasoning model from DeepSeek. Included as an additional check for whether larger models might perform better.

### 3.2.3 Request processing

Each proposal was submitted five times to each LLM and the average score used for that LLM. Previous research has shown that averaging multiple iterations increases correlations with human judgments (e.g., Thelwall & Yang, 2025), presumably by leveraging more of the LLM's internal knowledge as it uses its partially random mechanism to understand the task and produce its report.

The reports adhered to the requested structure, and a simple Python program was written to extract the four scores and take the proposal average of the five iterations for each score. This resulted in a set of four scores per document and LLM (two per proposal, one for the title/summary and one for the full proposal).

## *3.3 Analysis*

From the perspective of the utility of the information, the most important aspect of the LLM scores is their ranking rather than their exact values. This is because LLMs typically give high scores for tasks rather than carefully following the descriptors of the points in the scale. Despite these scores being essentially mostly wrong, the rankings are still often useful (Thelwall & Mohammadi, 2026). Thus, the primary test of utility of the LLM scores was the extent to which their rankings of the proposals agree with the expert rankings. This also aligns with the core funding task of essentially selecting the top ranked proposals for funding, stopping when the funding runs out. Normally, committees would often make minor adjustments to their decisions and not follow the rankings precisely. But this was not the case in the Swedish MRC. The original score set by each reviewer to a proposal was not changed in the process. Thus, Spearman rank correlations were used to assess the usefulness of the LLM predictions. Bootstrapped 95% confidence intervals were calculated for an indication of statistical significance.

A different goal that needs a different statistic is assessing the extent of underlying agreement on scoring for a task. There are many different measures for this. To assess the agreement between individual reviewers and individual LLMs, the intraclass correlation (ICC) metric would be the most appropriate. Specifically, ICC (1,1) is suitable when the two reviewer scores are from multiple reviewers. This is like the Pearson correlation but does not assume the same mean and variance for reviewer scores (Koo & Li, 2016; Shrout & Fleiss, 1979). Nevertheless, it does not work here because it relies on the reviewers working on the same scale range, which LLMs do not – they tend to focus on one high score. Thus, the only useful measure is for rank order, for which the Spearman correlation is appropriate.

## 4 Results

The LLMs had weak but usually positive and statistically significant positive correlations with mean reviewer scores for all dimensions and overall, both for the title and summary and the title, summary and main text (Figures 2-4). Although the sample sizes are too small for robust comparisons, the LLM predictions seem worst for the Mobility panel. Of the models, Gemma 3 27b had the strongest correlations, although the differences between models are mostly not statistically significant. Neither the title and summary nor the full text are clearly superior overall: it does not seem to matter which is fed to any of the LLMs. Moreover, averaging across models seems to improve the correlation, although the difference is small and not statistically significant in any case.

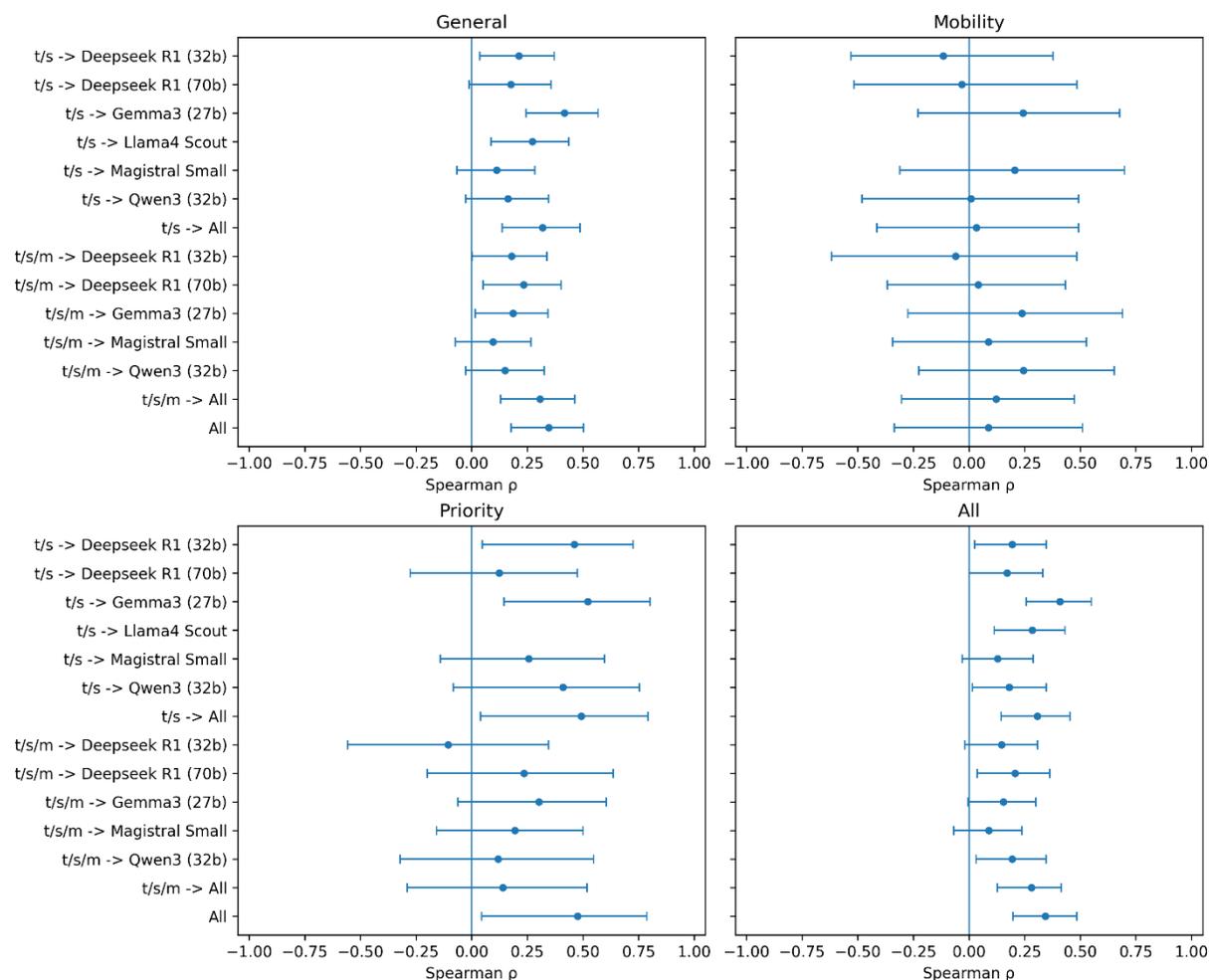

Figure 1. Spearman correlations between LLM scores (average of five) and mean reviewer scores for **research questions** by panel and overall based on proposal titles and summaries (t/s) or titles, summaries and main texts (t/s/m). "All" averages across all LLM scores, including those based on proposal titles and summaries. Error bars indicate 95% confidence intervals. Llama Scout is missing from the titles, summaries and main texts set because it produced few valid scores.

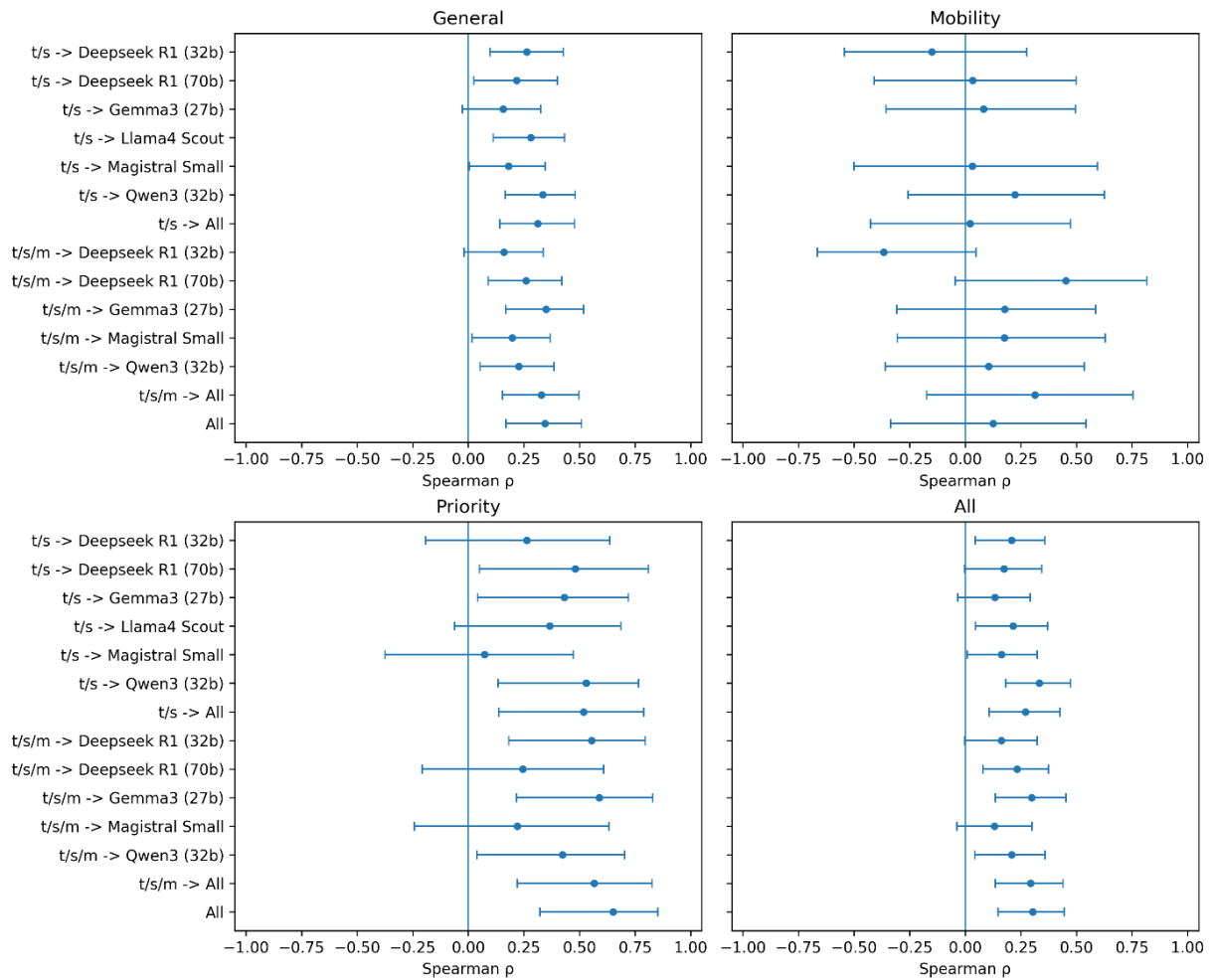

Figure 2. Spearman correlations between LLM scores (average of five) and mean reviewer scores for **methods** by panel and overall based on proposal titles and summaries (t/s) or titles, summaries and main texts (t/s/m). "All" averages across all LLM scores, including those based on proposal titles and summaries. Error bars indicate 95% confidence intervals. Llama Scout is missing from the titles, summaries and main texts set because it produced few valid scores.

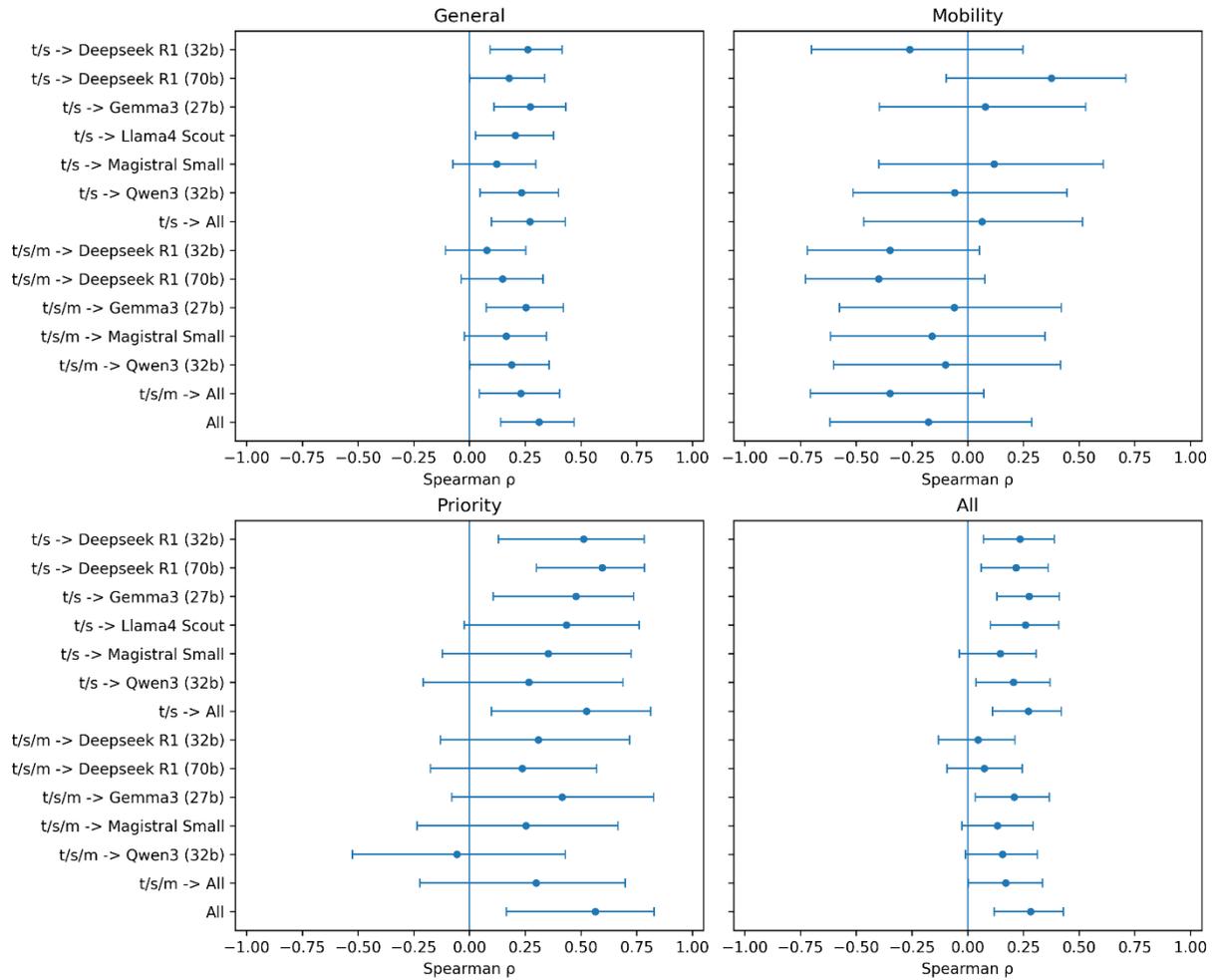

Figure 3. Spearman correlations between LLM scores (average of five) and mean reviewer scores for **applicant competence** by panel and overall based on proposal titles and summaries (t/s) or titles, summaries and main texts (t/s/m). "All" averages across all LLM scores, including those based on proposal titles and summaries. Error bars indicate 95% confidence intervals. Llama Scout is missing from the titles, summaries and main texts set because it produced few valid scores.

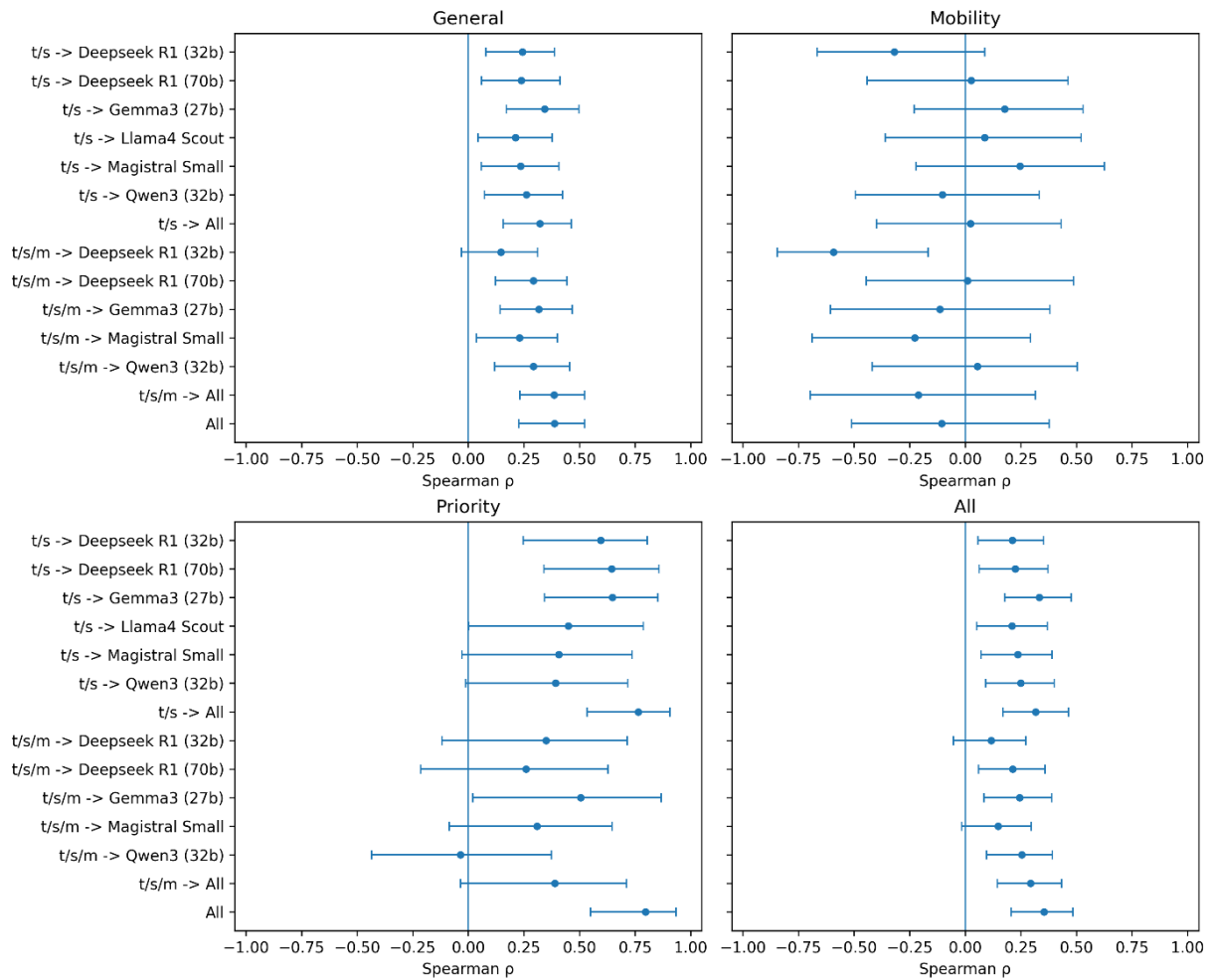

Figure 4. Spearman correlations between LLM scores (average of five) and mean reviewer scores **overall** by panel and overall based on proposal titles and summaries (t/s) or titles, summaries and main texts (t/s/m). "All" averages across all LLM scores, including those based on proposal titles and summaries. Error bars indicate 95% confidence intervals. Llama Scout is missing from the titles, summaries and main texts set because it produced few valid scores.

In 26 cases there were two panel scores for the same proposal. For these cases, Spearman correlations between panel scores were calculated, giving the following results.
- General: 0.63 (0.26, 0.83), n=21.
- Mobility: 0.83 (-1.00, 1.00), n=4.
- Priority: 0.35 (-0.54, 0.94), n=9.
- All: 0.62 (0.32, 0.81), n=26.

Thus, where the sample sizes are large enough for a reasonable test (General and All), the panels agreed about twice as strongly with each other as the LLMs agree with their average scores. Of course, the agreement rate between individual reviewers is probably less than the agreement rate between panels, so the LLM agreement rate may be closer to that of individual experts.

The LLMs agreed with each other more than with the human reviewers overall. For example, for the overall scores, the average correlation between LLMs was 0.34 whereas their average correlation with the reviewer scores was 0.22, The rank agreement between LLMs was highest when the input was the title and summary (Figure 5). Thus, they seem to process the text in broadly similar ways. The higher correlations obtained by averaging LLMs suggests

13that they may extract different relevant patterns from the proposals, with the averaging exploiting this. The lower agreement on longer texts may reflect greater scope for algorithm variations to leverage information differently with more text.

| | t/s -> Deepseek R1 (32b) | t/s -> Deepseek R1 (70b) | t/s -> Gemma3 (27b) | t/s -> Llama4 Scout | t/s -> Magistral Small | t/s -> Qwen3 (32b) | t/s -> All | t/s/m -> Deepseek R1 (32b) | t/s/m -> Deepseek R1 (70b) | t/s/m -> Gemma3 (27b) | t/s/m -> Magistral Small | t/s/m -> Qwen3 (32b) | t/s/m -> All | All | Expert |
|---|---|---|---|---|---|---|---|---|---|---|---|---|---|---|---|
| t/s -> Deepseek R1 (32b) | 1.00 | 0.48 | 0.61 | 0.34 | 0.30 | 0.59 | 0.76 | 0.26 | 0.30 | 0.30 | 0.26 | 0.27 | 0.41 | 0.70 | 0.21 |
| t/s -> Deepseek R1 (70b) | 0.48 | 1.00 | 0.64 | 0.48 | 0.41 | 0.56 | 0.82 | 0.31 | 0.34 | 0.17 | 0.26 | 0.26 | 0.42 | 0.75 | 0.23 |
| t/s -> Gemma3 (27b) | 0.61 | 0.64 | 1.00 | 0.50 | 0.40 | 0.64 | 0.86 | 0.19 | 0.32 | 0.32 | 0.26 | 0.28 | 0.42 | 0.78 | 0.33 |
| t/s -> Llama4 Scout | 0.34 | 0.48 | 0.50 | 1.00 | 0.25 | 0.43 | 0.57 | 0.12 | 0.12 | 0.19 | 0.29 | 0.10 | 0.22 | 0.49 | 0.21 |
| t/s -> Magistral Small | 0.30 | 0.41 | 0.40 | 0.25 | 1.00 | 0.36 | 0.55 | 0.03 | 0.23 | 0.11 | 0.14 | 0.26 | 0.25 | 0.48 | 0.24 |
| t/s -> Qwen3 (32b) | 0.59 | 0.56 | 0.64 | 0.43 | 0.36 | 1.00 | 0.81 | 0.16 | 0.35 | 0.16 | 0.19 | 0.29 | 0.38 | 0.70 | 0.25 |
| t/s -> All | 0.76 | 0.82 | 0.86 | 0.57 | 0.55 | 0.81 | 1.00 | 0.27 | 0.38 | 0.27 | 0.29 | 0.34 | 0.48 | 0.89 | 0.32 |
| t/s/m -> Deepseek R1 (32b) | 0.26 | 0.31 | 0.19 | 0.12 | 0.03 | 0.16 | 0.27 | 1.00 | 0.27 | 0.38 | 0.14 | 0.28 | 0.62 | 0.48 | 0.12 |
| t/s/m -> Deepseek R1 (70b) | 0.30 | 0.34 | 0.32 | 0.12 | 0.23 | 0.35 | 0.38 | 0.27 | 1.00 | 0.34 | 0.28 | 0.39 | 0.74 | 0.62 | 0.21 |
| t/s/m -> Gemma3 (27b) | 0.30 | 0.17 | 0.32 | 0.19 | 0.11 | 0.16 | 0.27 | 0.38 | 0.34 | 1.00 | 0.45 | 0.39 | 0.71 | 0.53 | 0.24 |
| t/s/m -> Magistral Small | 0.26 | 0.26 | 0.26 | 0.29 | 0.14 | 0.19 | 0.29 | 0.14 | 0.28 | 0.45 | 1.00 | 0.22 | 0.54 | 0.48 | 0.15 |
| t/s/m -> Qwen3 (32b) | 0.27 | 0.26 | 0.28 | 0.10 | 0.26 | 0.29 | 0.34 | 0.28 | 0.39 | 0.39 | 0.22 | 1.00 | 0.67 | 0.55 | 0.25 |
| t/s/m -> All | 0.41 | 0.42 | 0.42 | 0.22 | 0.25 | 0.38 | 0.48 | 0.62 | 0.74 | 0.71 | 0.54 | 0.67 | 1.00 | 0.80 | 0.29 |
| All | 0.70 | 0.75 | 0.78 | 0.49 | 0.48 | 0.70 | 0.89 | 0.48 | 0.62 | 0.53 | 0.48 | 0.55 | 0.80 | 1.00 | 0.35 |
| Expert | 0.21 | 0.23 | 0.33 | 0.21 | 0.24 | 0.25 | 0.32 | 0.12 | 0.21 | 0.24 | 0.15 | 0.25 | 0.29 | 0.35 | 1.00 |

Figure 5. Spearman correlations between LLM scores (average of five) and mean reviewer scores overall. Llama Scout is missing from the title, summary and main text set because it produced few valid scores.

## 5 Discussion

This study has very many limitations, most of which are likely to have reduced the strength of the LLM correlations with expert judgment relative to a perfect test. The main limitations are that the data is old, and the sample sizes are too small to analyse the panels separately, which would be the most robust approach. In addition, the LLMs would not have taken into account host environment quality and reputation, and the human reviewers probably did. Another important limitation that may have increased the correlation is that the outcomes of the funding process is partly known and may be in the training data, in the sense that funded researchers may acknowledge their funding on their CVs and in their publications. This creates a risk that the LLMs may be cheating by knowing the outcome, but this seems unlikely because they do not seem to know details about individual academic documents. For example, they rarely answer basic questions about published journal articles, even if they are highly cited (Thelwall, 2026). It is also not clear that the criteria and application processes from 1994 are still relevant today. Moreover, the translation step for the Swedish applications may have influenced the scoring.

4In comparison to prior research, whilst no study has used LLMs to predict grant review scores before, the findings add to the growing body of evidence for multiple document types that LLMs have a capability to produce scores that correlate positively with expert scores (Kousha et al., 2025; Thelwall, 2025; Thelwall & Cox, 2025; Thelwall & Yaghi, 2025a). The current correlations are weaker than most previously found, suggesting that this task may be relatively difficult for LLMs. A possible explanation is that grant review is intrinsically difficult, although the inter-human correlation was much higher (albeit with wide confidence intervals) so this does not seem likely. Alternatively, the age of the proposals might have made the task more difficult. Another explanation is that the proposal evaluation is highly person-centred, for fellowships, and for a relatively small group, such as Swedish medical researchers, the human reviewers could have taken more tacit and background knowledge into account when evaluating the individuals.

## 6 Conclusions

The results show for the first time, albeit on a limited scale, that LLMs have a weak but statistically significant ability to score grant proposals in a way that aligns with human scores in rank, although not in magnitude. This applies to medium sized LLMs, with Gemma 3 27b being possibly the best model available at the start of 2026, although not all have been tested and differences between models were not statistically significant. The predictions can be made almost equally well based on a proposal's title and abstract, without its main text, and so are clearly pattern-based guesses rather than "evaluations" in any meaningful sense. Moreover, averaging the scores from multiple models and both types of input (i.e., titles and summaries with and without main texts) seems to give the best results, with correlations that are 56% of the correlations between the expert panels.

The results from the current study broadly support La Caixa's existing use of LLMs for application triage, with humans making the final decision. Another possible application is in the tie-breaking stage when only a subset of multiple applications with similar scores can be funded. Currently, this decision may be made by committees after reading the proposals and it seems possible that it would be at least as fair, and quicker, to allow LLM scores to be the tiebreaker. If this is not considered acceptable, then they might at least provide an initial ranking to guide the final decision. Of course, LLMs can also be used as an addition to the human scores to cross-check for bias or "unlucky" applications with multiple harsh human reviewers or for "lucky" applications with multiple generous reviewers.

## 7 Acknowledgement

This study was partly funded by ESRC Metascience grant UKRI1486.